\documentclass{ws-p10x7}
\newcommand{\lesssim}%
{\mathop{\raisebox{-.4ex}{\rlap{$\sim$}} \raisebox{.4ex}{$<$}}}

\renewcommand{\cropmarks}{}
\begin{document}

\title{$B$ and $D$ Mesons in Lattice QCD \hfill {\rm FERMILAB-CONF-00/256-T}}

\author{Andreas S. Kronfeld}

\address{Theoretical Physics Department,
Fermi National Accelerator Laboratory,
Batavia, IL, USA}

\twocolumn[\maketitle\abstract{Computational and theoretical
developments in lattice QCD calculations of $B$ and $D$ mesons
are surveyed.
Several topical examples are given: new ideas for calculating the HQET 
parameters $\bar{\Lambda}$ and $\lambda_1$; form factors needed to 
determine $|V_{cb}|$ and $|V_{ub}|$; bag parameters for the mass 
differences of the $B$ mesons; and decay constants.
Prospects for removing the quenched approximation are discussed.}
\thispagestyle{plain}]

\section{Introduction}
\label{sec:intro}

\begin{figure}[b]
\vskip -12pt
	\includegraphics[width=0.48\textwidth]{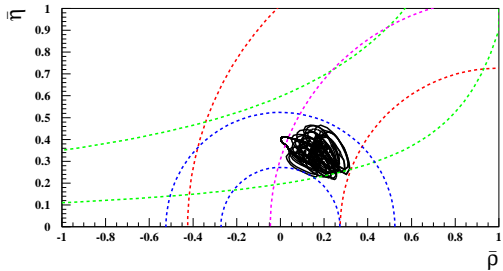}
\vskip -6pt
	\caption[fig:utps]{Constraints on the unitarity triangle from $CP$
	conserving $B$ decays and indirect $CP$ violation in the kaon.
	From Plaszczynski and Schune.\cite{Plaszczynski:1999xs}}
	\label{fig:utps}
\end{figure}

In the standard model, interactions involving the
Cabibbo-Kobayashi-Maskawa (CKM) matrix violate $CP$, with strength
proportional to the area of the ``unitarity triangle.''
Fig.~\ref{fig:utps} shows a recent summary\cite{Plaszczynski:1999xs} of 
the triangle.
The dominant uncertainties are theoretical, coming from 
non-perturbative QCD.
Each blob shows experimental uncertainties for fixed theoretical
inputs, so the range of blobs illustrates the theoretical uncertainties.
If measuring the apex $(\bar{\rho},\bar{\eta})$ were the only goal, one
might conclude from Fig.~\ref{fig:utps} that the most pressing issue is
to reduce the theoretical uncertainties,
which would require greater investment in computing for lattice QCD.

Measuring $(\bar{\rho},\bar{\eta})$ is not the most exciting goal,
however.
``High-energy physics is exciting and will remain
exciting, precisely because it exists in a state of permanent
revolution.''\@\cite{Lykken:1999gn}
That means we would prefer to discover additional, non-KM sources of
$CP$ violation.
Indeed,
``it is possible, likely, unavoidable, that the standard
model's picture of $CP$ violation is incomplete.''\cite{Nir}

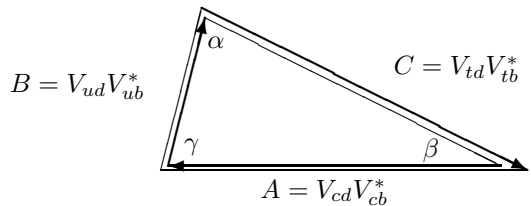
\begin{figure}[b]
\vskip -12pt
\setlength{\unitlength}{0.5pt}
\begin{picture}(336,120)(-120,0)

\thicklines
\put(0,0){\vector(1,4){28}}
\put(70,-24){$A=V_{cd}V_{cb}^*$}
\put(252,0){\vector(-1,0){252}}
\put(-120,56){$B=V_{ud}V_{ub}^*$}
\put(12,12){$\gamma$}

\thinlines
\put(28,112){\line(2,-1){224}}

\setlength{\unitlength}{0.55pt}
\thicklines
\put(-5,-3){\begin{picture}(336,120)
\put(28,112){\vector(2,-1){224}}
\put(160,66){$C=V_{td}V_{tb}^*$}
\put(32,84){$\alpha$}
\put(180,9){$\beta$}
\end{picture}
}

\thinlines
\put(-5,-3){\begin{picture}(336,120)
\put(0,0){\line(1,4){28}}
\put(252,0){\line(-1,0){252}}
\end{picture}
}

\end{picture}
\vskip 12pt
\caption[fig:2t]{Two different unitarity triangles:
the tree   triangle $B\gamma A$, and
the mixing triangle $\alpha C\beta$.}
\label{fig:2t}
\end{figure}

Lattice QCD can aid the discovery of new sources of~$CP$ violation
and may be essential.
A lot of information will be necessary to figure out what is going on 
at short distances.
One way to think about this is sketched in Fig.~\ref{fig:2t}.
The triangle $B\gamma A$ is determined from (quark-level) tree
processes.
The side $B$ requires
$|V_{ud}|$ from $n\to pe^-\bar{\nu}$, and
$|V_{ub}|$ from $B^-\to \rho^0l^-\bar{\nu}$ or
$\bar{B}^0\to\pi^+l^-\bar{\nu}$;
the angle $\gamma$ requires the $CP$ asymmetry of 
$B^{\pm}\to D^0_{CP} K^\pm$ (or $B_s\to D_s^\pm K^\mp$);
the side $A$ requires $|V_{cd}|$ from $D^0\to\pi^-l^+\nu$, and
$|V_{cb}|$ from $B^-\to D^{0(*)}l^-\bar{\nu}$.
One could call this the ``tree triangle''.
The triangle $\alpha C\beta$ is determined from mixing processes
(including interference of decays with and without mixing).
The angle $\alpha$ requires the asymmetry of $B\to\rho\pi$;
the side $C$ requires
$|V_{td}|$ from $\Delta m_{B_d^0}^{ }$, and
$|V_{tb}|$ from $t\to W^+b$;
the angle $\beta$ requires the asymmetry of $B^0\to J/\psi K_S$.
One could call this the ``mixing triangle''.

Checking whether the mixing triangle agrees with the tree triangle
tests for new physics in the amplitude of
$B^0_d$-$\bar{B}^0_d$ mixing.
New physics in the magnitude muddles the extraction of $|V_{td}|$, and
new physics in the phase     muddles the extraction of angles~$\alpha$
and~$\beta$.
Similarly, taking $\Delta m_{B_s^0}^{ }$,
$B_s\to D_s^\pm K^\mp$, and $B_s\to J/\psi\,\eta^{(\prime)}$
(or $J/\psi\,\phi$)
sorts out new physics in $B^0_s$-$\bar{B}^0_s$ mixing.

These tests are impossible without knowing the sides accurately,
so hadronic matrix elements are needed.
In a few cases a symmetry provides it, e.g., isospin cleanly
yields the matrix element for $n\to pe^-\bar{\nu}$.
For the others, we must ``solve'' non-perturbative QCD and, therefore,
\emph{you} need lattice calculations.

You are probably tired of waiting and may ask why results should come
any time soon.
The fastest of today's computers are now powerful enough to eliminate
the sorest point: the quenched approximation.
Also, (lattice) theorists have slowly developed a culture of 
estimating systematic uncertainties, which is now not bad and would 
improve if more non-practitioners became sufficiently informed about 
the methods to make constructive suggestions.

The rest of this talk starts with some theoretical aspects that 
might make it easier for the outsider to judge the systematic errors 
of heavy quarks on the lattice.
Then I~show recent results needed for the sides $A$, $B$, and~$C$,.

\section{Lattice Spacing Effects (Theory)}
\label{sec:theory}

Lattice QCD calculates matrix elements by computing the functional
integral, using a Monte Carlo with importance sampling.
Hence, there are statistical errors.
This part of the method is well understood and, these days, rarely
leads to controversy.
When conflicts do arise, they usually originate in the treatment
of systematics.
The non-expert does not need to know how the Monte Carlo works, but 
can develop some intuition of how the systematics work.
Don't be put off by lattice jargon: the main tool is familiar to all: 
it is effective field theory.

Lattice spacing effects can be cataloged with Symanzik's local
effective Lagrangian (LE${\cal L}$).\@\cite{Symanzik:1979ph}
Finite-volume effects can be controled and exploited with a general,
massive quantum field theory.\@\cite{Luscher:1986dn}
The computer algorithms work better for the strange quark than for
down or up, but the dependence on $m_q$ can be understood and
controlled via the chiral Lagrangian.\@\cite{Gasser:1984yg}
Finally, discretization effects of the heavy-quark mass $m_Q$ are
treated with HQET\cite{Neubert:1994mb} or NRQCD.\@\cite{Caswell:1986ui}
In each case one can control the extrapolation of artificial, 
numerical data, if one generates numerical data close enough to the
real world.

Volume effects are unimportant in what follows, and chiral 
perturbation theory is a relatively well-known subject.
Therefore, here I~will focus on the effective field theories that help 
us control discretization effects.

\subsection{Symanzik's LE${\cal L}$}

Symanzik's formalism\cite{Symanzik:1979ph} describes the 
lattice theory with continuum~QCD:
\begin{equation}
	{\cal L}_{\rm lat} \doteq {\cal L}_{\rm cont} +
		\sum_i a^{s_i} C_i(a;\mu) {\cal O}_i(\mu),
\end{equation}
where the symbol $\doteq$ means ``has the same physics as''.
The LE${\cal L}$ on the right-hand side is defined in, say, the
$\overline{\rm MS}$ scheme at scale $\mu$.
The coefficients $C_i$ describe short-distance physics, so they depend
on the lattice spacing~$a$.
The operators do not depend on~$a$.

If $\Lambda_{\rm QCD}a$ is small enough the higher terms
can be treated as perturbations.
So, the $a$ dependence of the proton mass is
\begin{equation}
	m_p(a) = m_p + a C_{\sigma F}
		\langle p|\bar{\psi}\sigma\cdot F\psi|p\rangle,
	\label{eq:mp}
\end{equation}
taking the leading operator for Wilson fermions as an example.
To reduce the second term one might try to reduce $a$ greatly, but CPU 
time goes as $a^{-\rm (5~or~6)}$.
It is more effective to combine several data sets and extrapolate, 
with Eq.~(\ref{eq:mp}) as a guide.
It is even better to adjust things so $C_{\sigma F}$ is 
$O(\alpha_s^\ell)$ or $O(a)$, which is called Symanzik improvement of 
the action.
For light hadrons, a combination of improvement and extrapolation is 
best, and you should look for both.

\subsection{HQET for large $m_Q$}

The Symanzik theory, as usually applied, assumes $m_qa\ll1$.
The bottom and charm quarks' masses in lattice units are at present
large: $m_ba\sim1$--2 and $m_ca$ about a third of that.
It will not be possible to reduce $a$ enough to make $m_ba\ll1$
for many, many years.
So, other methods are needed to control the lattice spacing
effects of heavy quarks.
There are several alternatives:
\begin{enumerate}
\item static approximation\cite{Eichten:1990zv}
\item lattice NRQCD\cite{Lepage:1987gg}
\item extrapolation from $m_Q\approx m_c$ up to $m_b$
\item[$3'\!.$] combine 3 with 1
\item normalize systematically to HQET\cite{El-Khadra:1997mp}
\end{enumerate}
All use HQET in some way.
The first two discretize continuum HQET;
method~1 stops at the leading term, and method~2 carries the
heavy-quark expansion out to the desired order.
Methods~3 and~$3'$ keep the heavy quark mass artificially small and
appeal to the $1/m_Q$ expansion to extrapolate back up to $m_b$.
Method~4 uses the same lattice action as method~3, but uses the
heavy-quark expansion to normalize and improve it.
Methods~2 and~4 are able to calculate matrix elements directly at
the $b$-quark mass.

The methods can be compared and contrasted by \emph{describing} the 
lattice theories with HQET.\@\cite{Kronfeld:2000ck}
This is, in a sense, the opposite of \emph{discretizing} HQET.
One writes down a (continuum) effective Lagrangian
\begin{equation}
	{\cal L}_{\rm lat} \doteq
		\sum_n {\cal C}^{(n)}_{\rm lat}(m_Qa; \mu) 
			{\cal O}^{(n)}_{\rm HQET}(\mu),
	\label{eq:hqet}
\end{equation}
with the operators defined exactly as in continuum HQET, so they do
not depend on $m_Q$ or~$a$.
As long as $m_Q\gg\Lambda_{\rm QCD}$ this description makes
sense.
There are two short distances, $1/m_Q$ and the lattice spacing~$a$,
so the short-distance coefficients ${\cal C}^{(n)}_{\rm lat}$ depend 
on~$m_Qa$.
Since all dependence on $m_Qa$ is isolated into the coefficients,
this description shows that heavy-quark lattice artifacts arise only
from the mismatch of the ${\cal C}^{(n)}_{\rm lat}$ and their continuum 
analogs~${\cal C}^{(n)}_{\rm cont}$.

For methods~1 and~2, Eq.~(\ref{eq:hqet}) is just a
Symanzik LE${\cal L}$.
For lattice NRQCD we recover the well-known result that some of the 
coefficients have power-law divergences.\@\cite{Lepage:1987gg}
So, to take the continuum limit one must add more and more terms to
the action.
This leaves a systematic error, which, in practice, is usually
accounted for conservatively.

Eq.~(\ref{eq:hqet}) is more illuminating for methods~3 and~4, which 
use Wilson fermions (with an improved action).
Wilson fermions have the same degrees of freedom and heavy-quark
symmetries as continuum QCD, so the HQET description is admissible for
all~$m_Qa$.
Method~4 matches the coefficients of Eq.~(\ref{eq:hqet}) term by term, 
by adjusting the lattice action.
In practice, this is possible only to finite order, so there are
errors $({\cal C}^{(n)}_{\rm lat}-{\cal C}^{(n)}_{\rm cont})%
\langle{\cal O}^{(n)}_{\rm HQET}\rangle$, starting with some $n$.
Method~3 reduces $m_Qa$ until the mismatch is of order $(m_Qa)^2\ll 1$
(or $\lesssim 1$).
This runs the risk of reducing $m_Q$ until the heavy-quark 
expansion falls apart.

The non-expert can get a feel for which methods are most appropriate by
asking himself what order in $\Lambda_{\rm QCD}/m_b$ is needed.
For zeroth order, method~1 will do.
For the first few orders, the others are needed, although with
method~3 one should check that the calculation's
$\Lambda_{\rm QCD}/m_Q$ is small enough too.

\section{New Results}
\label{sec:results}

\subsection{$\bar{\Lambda}$ and $\lambda_1$}

The matching of lattice gauge theory to HQET provides a new way to 
calculate matrix elements of the heavy-quark 
expansion.\@\cite{Kronfeld:2000gk}
The spin-averaged $B^*$-$B$ mass is given by\cite{Falk:1993wt}
\begin{equation}
	\bar{M} = m + \bar{\Lambda} - \lambda_1/2m,
	\label{eq:M}
\end{equation}
where $m$ is the heavy quark mass,
and $\bar{M}=\frac{1}{4}(3M_{B^*}+M_{B})$.
The lattice changes the short-distance definition of the
quark mass:\cite{Kronfeld:2000ck}
\begin{equation}
	\bar{M}_1 - m_1 = \bar{\Lambda}_{\rm lat} -
	{\lambda_1}_{\rm lat}/2m_2.
	\label{eq:M1}
\end{equation}
Because the lattice breaks Lorentz symmetry, $m_1\neq m_2$, but they 
are still calculable in perturbation theory.\@\cite{Mertens:1998wx}
The lambdas in Eq.~(\ref{eq:M1}) are labeled ``lat'' because
they suffer lattice artifacts from the gluons and light quark.

After fitting a wide range of lattice data to Eq.~(\ref{eq:M1}) 
\emph{and} taking the continuum limit, we find\cite{Kronfeld:2000gk}
$\bar{\Lambda} = 0.68^{+0.02}_{-0.12}~{\rm GeV}$ and
$\lambda_1 = -(0.45\pm 0.12)~{\rm GeV}^2$
in the quenched approximation.
The lambdas appear also in the heavy-quark expansion of inclusive decays.
Although the current analysis is thorough, there are several ways to
improve it.\@\cite{Kronfeld:2000gk}
For example, $\bar{\Lambda}_{\rm lat}$ has an unexpectedly large
$a$~dependence, so the analysis should be repeated with an action
for which $C_{\sigma F}$ in Eq.~(\ref{eq:mp}) is~$O(a)$.

\subsection{$B\to\pi l\nu$ form factors and $V_{ub}$}

It is timely to discuss $\bar{B}^0\to\pi^+l^-\bar{\nu}$,
because there are three calculations to compare,
using lattice NRQCD (method~2),\cite{Aoki:2000ij}
the extrapolation method (method~3),\cite{Bowler:2000xn}
and the HQET matching method (method~4).\@\cite{Ryan:2000kx}
UKQCD's work is final,\cite{Bowler:2000xn}
and the other two are preliminary.\@\cite{Aoki:2000ij,Ryan:2000kx}
The decay rate requires a form factor, called~$f_+(E)$, which depends
on the pion's energy in the $B$'s rest frame, $E=v\cdot p_\pi$.
It is related to the matrix element $\langle\pi|V^\mu|B\rangle$, which
can be computed in lattice QCD.
The systematics are smallest when the pion's three-momentum is small.

\begin{figure}[b]
    \includegraphics[width=0.48\textwidth]{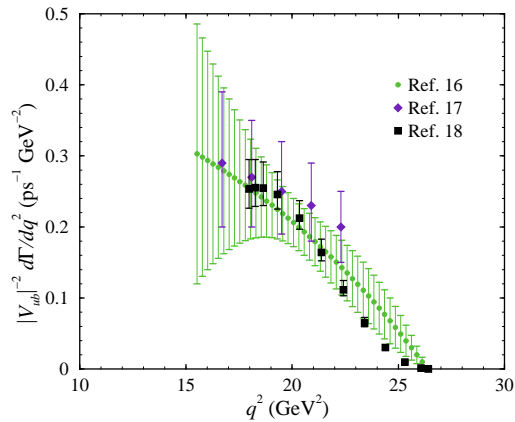}
	\caption[fig:b2pi]{Recent results for the decay $B\to \pi l\nu$.}
	\label{fig:b2pi}
\end{figure}

The three recent results are compared in Fig.~\ref{fig:b2pi}.
The error bars shown are statistical only.
For NRQCD these are larger than expected.\@\cite{Aoki:2000ij}
For the other two, the comparison gives a fair idea of systematics 
that are not common to both, because
the extrapolation of heavy quark mass needed with
method~3 amplifies the statistical error.\@\cite{Bowler:2000xn}
The other two works\cite{Aoki:2000ij,Ryan:2000kx}
compute directly at the $b$ quark mass and, thus, circumvent this
problem.
The heavy quark masses of UKQCD\cite{Bowler:2000xn} are all below
1.3~GeV, and as low as 500~MeV, so one might worry whether the
heavy-quark expansion applies.

\subsection{$B^-\to D^{0(*)}l^-\bar{\nu}$ and $V_{cb}$}

The form factors ${\cal F}_{B\to D^{(*)}}$ of the decays
$B\to D^{(*)}l\nu$ are normalized to unity for infinite quark masses.
What is needed from lattice QCD, therefore, is the deviation from
the unity for physical quark masses.
Hashimoto \emph{et al.}\cite{Hashimoto:2000yp,Simone:2000nv} have
devised methods based on double ratios, in which all the uncertainties
cancel in the symmetry limit.
Consequently, all errors scale as ${\cal F}-1$, not as ${\cal F}$.

For $B\to Dl\nu$ they find\cite{Hashimoto:2000yp} (published)
\begin{equation}
	\mathcal{F}_{B\to D}(1) = 
		1.058 \pm 0.016 \pm 0.003^{+0.014}_{-0.005},
	\label{eq:D}
\end{equation}
where error bars are from statistics, adjusting the quark masses, 
and higher-order radiative corrections.
For $B\to D^*l\nu$ they find\cite{Simone:2000nv} (still preliminary)
\begin{equation}
	\begin{array}{r@{\,\pm\,}l}
	\mathcal{F}_{B\to D^*}(1) = 
		0.935 & 0.022^{+0.008}_{-0.011} \\ & 0.008 \pm 0.020,
	\end{array}
	\label{eq:Dstar}
\end{equation}
where now the last uncertainty is from $1/m_Q^3$.
In both results, an ongoing test of the lattice spacing dependence is
not included, but that will probably not be noticeable.
More seriously, these results are, once again, in the quenched
approximation, but the associated uncertainty are still only
a fraction of ${\cal F}-1$.
Both results will be updated soon, with calculations at a second
lattice spacing and refinements in the radiative corrections.

\subsection{$B^0_q$-$\bar{B}^0_q$ mixing: $B_B$, $f_B$, and $V_{tq}$}
The mass difference of $CP$ eigenstates is
\begin{equation}
	\Delta m_{B^0_q} = \frac{G_F^2m_W^2S_0}{16\pi^2}
	|V_{tq}^*V_{tb}|^2
	\eta_B \langle Q_q^{\Delta B=2}\rangle,
	\label{eq:BB}
\end{equation}
where the light quark $q$ is $d$ or~$s$, $S_0$ is an Inami-Lim 
function, and
$\langle Q_q^{\Delta B=2}\rangle$ is
\begin{equation}
	\langle \bar{B}_q^0| Q_q^{\Delta B=2}|B_q^0\rangle =
	{\textstyle\frac{8}{3}}m^2_{B_q}f^2_{B_q} B_{B_q} .
	\label{eq:fBBB}
\end{equation}
The $\mu$ dependence in $\eta_B$ and $Q_q^{\Delta B=2}$ cancels.
New physics could compete with the $W$ and $t$ box 
diagrams and change Eq.~(\ref{eq:BB}).

Lattice QCD gives matrix elements, so the basic results are
$\langle Q_q^{\Delta B=2}\rangle$ and~$f_{B_q}$.
It is often stated that uncertainties in
$B_{B_q}$ and $\xi^2=f_{B_s}^2B_{B_s}/f_{B_s}^2B_{B_s}$
should be small, because they are ratios.
Some cancelation should occur, but only if one can show
that the errors are under control.
At present there are unresolved issues in method~3,
so one should be cautious.

With that warning, results for $B_B$ from three groups are in
Table~\ref{tab:BB}.
\begin{table}[t]
\begin{tabular}{ccl}
\hline \hline
group   & method      & 
\multicolumn{1}{c}{$B_{B_d}$(4.8~GeV)} \\
\hline
JLQCD\cite{Hashimoto:1999ck} & 2 & $0.85 \pm 0.03 \pm 0.11$ \\
  APE\cite{Becirevic:2000nv} & 3 & $0.93 \pm 0.08^{+00}_{-06}$ \\
UKQCD\cite{Lellouch:1999ym}  & 3 & $0.92 \pm 0.04^{+03}_{-00}$ \\
\hline \hline
\end{tabular}
\caption[tab:BB]{Recent quenched results for $B_B$.}
\label{tab:BB}
\end{table}
Note that JLQCD now includes the short-distance part of the $1/m_Q$
contribution.\@\cite{Hashimoto:2000eh}
APE\cite{Becirevic:2000nv} (final) and UKQCD\cite{Lellouch:1999ym}
(preliminary) both extrapolate linearly in $1/m_Q$
from charm (e.g., $1.75~{\rm GeV}<m_{``B''}<2.26~{\rm GeV}$ for APE).
It is not clear whether the first term of the heavy-quark expansion is
adequate here; everyone working in $B$ physics can and should
form his or her own opinion.
It is also not clear how the $(m_Qa)^2$ lattice artifacts of method~3
fare through the $1/m_Q$ extrapolation.
It is likely that the systematic error is not well controlled and,
thus, possibly underestimated in for the last two rows
Table~\ref{tab:BB}.
At present one should prefer the JLQCD results.


The MILC\cite{Bernard:2000nv} and CP-PACS\cite{AliKhan:2000uq} groups
have new, \emph{preliminary} unquenched calculations of the heavy-light
decay constants $f_B$, $f_{B_s}$, $f_D$, and~$f_{D_s}$.
Both use method~4.
Both have results at several lattice spacings,
so they can study the continuum limit.
The status for Osaka is tabulated in Table~\ref{tab:fB}.
\begin{table}[t]
\begin{tabular}{rll}
\hline \hline
          $f_P(n_f)$  &
 \multicolumn{1}{c}{MILC\cite{Bernard:2000nv}}   &
 \multicolumn{1}{c}{CP-PACS\cite{AliKhan:2000uq}} \\
\hline
$f_B(0)$ &
	$171\pm6\pm17^{+21}_{-~4}$ & $190 \pm 3 \pm 9$ \\
	      (2) &
	$190\pm6^{+20}_{-15}{}^{+9}_{-0}$ & $215 \pm 11 \pm 11$ \\
$f_{B_s}(0)$ &
	$197\pm5\pm23^{+25}_{-~6}$   & $224 \pm 2 \pm 15$ \\ 
	      (2) &
	$218\pm5^{+26}_{-23}{}^{+11}_{-0}$ & $250 \pm 10 \pm 13$ \\ 
$f_D(0)$ &
	$199\pm6\pm12^{+14}_{-~0}$ & $224 \pm 2 \pm 15$ \\
	      (2) &
	$213\pm4^{+14}_{-13}{}^{+7}_{-0}$ & $236 \pm 14 \pm 14$ \\
$f_{D_s}(0)$ &
	$222\pm5^{+19}_{-17}{}^{+15}_{-~0}$   & $252 \pm 1 \pm 18$ \\ 
	      (2) &
	$240\pm4^{+25}_{-23}{}^{+9}_{-0}$ & $275 \pm 10 \pm 17$ \\ 
\hline \hline
\end{tabular}
\caption[tab:fB]{Preliminary unquenched results for $f_B$, etc., in the
continuum limit.  All values in MeV.}
\label{tab:fB}
\end{table}
The first error is statistical, the second systematic.
MILC also provides an estimate of the error from quenching.
(With $n_f=2$ the strange quark is still quenched.)
CP-PACS\cite{AliKhan:2000uq} also has results with method~2,
which agree very well with method~4.
One should not, at this time, take the differences between the two
groups' central values very seriously.
It is more important to understand the different systematics of
methods~2, 3, and~4.

\section{Prospects}
\label{sec:prospects}

For $B$ physics it is important to remove the quenched approximation,
more so than to reduce the lattice spacing much further.
To do so, we need more computing.
Fermilab, MILC, and Cornell are building a cluster of PCs to tackle
the problem.\@\cite{pcqcd}
Our pilot cluster has 8 nodes with a Myrinet switch.
We plan to go up to 48--64 nodes, and then hope to assemble a cluster of
thousands of nodes.
The large cluster would evolve, by upgrading a third or so of the nodes
every year.
This is an ambitious plan, but not more ambitious than the experimental
effort to understand flavor mixing and $CP$ violation.

The last few years have seen significant strides in
understanding heavy quarks in lattice~QCD.
The progress has been both computational and theoretical, with
one guiding the other.
Calculations shown here, for $\bar{B}^0\to\pi^+l^-\bar{\nu}$,
$D^0\to\pi^-l^+\nu$, $B^-\to D^{0(*)}l^-\bar{\nu}$, and
$B_d^0$-$\bar{B}_d^0$ mixing are a subset, but in Fig.~\ref{fig:2t}
they are as basic as $A$, $B$, $C$.
With the right amount of support from the rest of the community, we hope
to obtain the tools needed to resolve the few outstanding problems and
to produce excellent unquenched results.
Indeed, the example of $f_B$ shows that this is already beginning.

\section*{Acknowledgments}
I~would like to thank Arifa Ali Khan, Claude Bernard, R\"udi
Burkhalter, Tetsuya Onogi, Hugh Shanahan and Akira Ukawa for
correspondence.
I~have benefited greatly from collaboration with Shoji Hashimoto,
Aida El-Khadra, Paul Mackenzie, Sin\'ead Ryan, and Jim Simone.
Fermilab is operated by Universities Research Association Inc.,
under contract with the U.S. Department of Energy.

\newpage
\begin{figure*}[ht]
\centering
\includegraphics[width=0.80\textwidth]{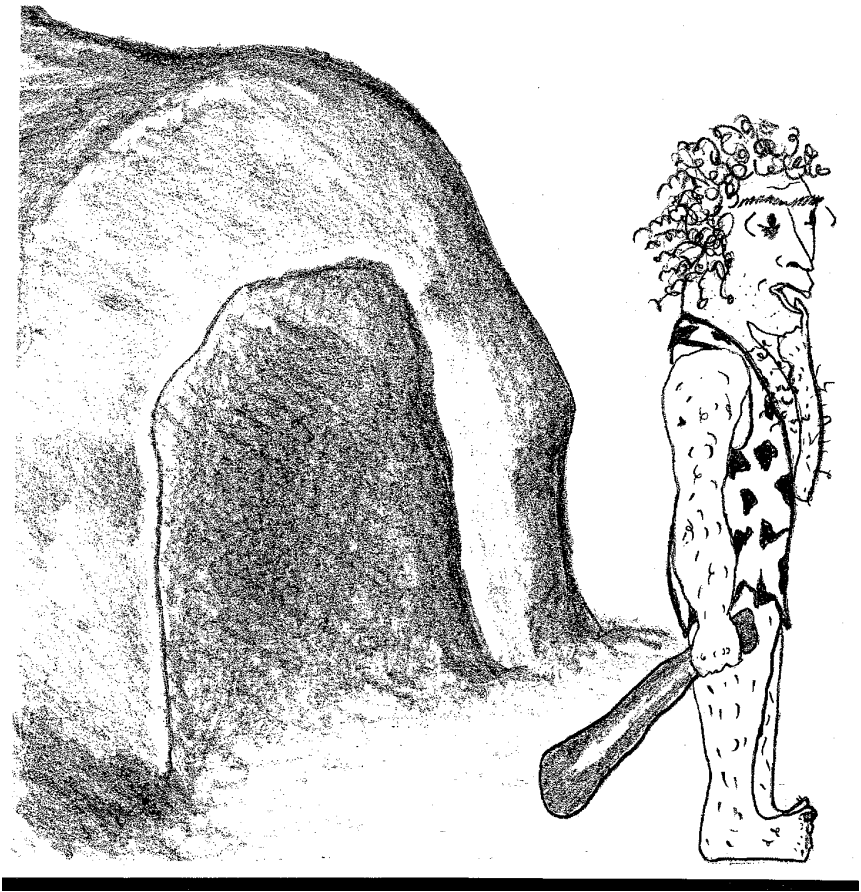}
\vskip 12pt
\caption[fig:caveman]{Artist's conception of the author 
striving to comprehend charm and beauty.
(\copyright 2000 Mercedes Kronfeld Jordan.)}
\end{figure*}

\end{document}